\documentclass[twoside]{article}
\usepackage{fbc,ebc}
\usepackage{cite}
\usepackage{epsfig}

\def\Rf#1#2#3#4{{#1} {#2} (19#3) #4}

\def\R2#1#2#3#4{{#1} {#2} (20#3) #4}

\def\PL{Phys. Lett.  {B}}   
\def\ZP{Z. Phys.  {C}}

\def\PRp{Phys. Reports}

\def\PRC{Phys. Rev. {C}}
\def\PRL{Phys. Rev. Lett. }

\def\JP{J. Phys. {G}}

\def\YaF{Yad. Fiz.}
\def\NC{Nuovo Cim.}

\def\ea{{et al.}}

\def\vs{\vspace*}
\def\hs{\hspace*}
\def\bea{\begin{eqnarray}}
\def\eea{\end{eqnarray}}
\def\be{\begin{equation}}
\def\ee{\end{equation}}
\def\bcr{\begin{center}}
\def\ecr{\end{center}}
\def\la{\label}
\def\ct{\cite}
\def\bi{\bibitem}

\def\mn{\rm min}
\def\mx{\rm max}
\def\rT{\rm T}
\def\rd{\rm d}
\def\e{\eta}

\def\ve{\varepsilon}
\def\vt{\vartheta}
\def\ef{\stackrel{\sim}{\e}}
\def\dep{\delta$${\ef}}
\def\ecf{\ef_{\rm 0}}   
\def\al{\langle}
\def\ar{\rangle}
\def\De{\Delta\e}

\def\dn{\delta n}
\def\r{\rho}
\def\ddf{\r(\ef)}
\def\se{\simeq}
\def\st{(\rm stat)}
\def\sy{(\rm syst)}

\title{Coherent particle production
                in collisions of relativistic nuclei \thanks{Invited talk
presented by
E.K.G. Sarkisyan at the 9th International Workshop on Multiparticle
Production,
Turin, Italy, June 12 - 17, 2000.}
}

\author{George L. Gogiberidze\address[JINR]{Joint Institute 
for Nuclear Research, RU-141980 Dubna, Moscow Region, 
Russia}\thanks{Now at Indiana Univ. Cyclotron Facility, Bloomington
IN 47408, USA}\thanks{On leave from Inst. of Physics, Tbilisi 380077,
Georgia}\thanks{Email address: goga@iucf.indiana.edu},
Edward K.G. Sarkisyan\address[TAU]{School of Physics and Astronomy,
The Raymond and Beverly Sackler Faculty of Exact Sciences,\\
Tel-Aviv University, IL-69978 Tel-Aviv, Israel}%
                      \thanks{Email address: edward@lep1.tau.ac.il}
and 
Liana K. Gelovani\addressmark[JINR]$^{\dagger\ddagger}$
}

\begin{document}


\begin{abstract}
 Here we give the results of our study of features of dense groups, or
spikes, of particles produced in Mg-Mg and C-Cu collisions at,
respectively, 4.3 and 4.5 GeV/$c$/nucleon aimed to search for a coherent,
{\v C}erenkov-like, mechanism of hadroproduction. 
 We investigate the distributions of spike centers and, for Mg-Mg
interactions, the energy spectra of negatively charged particles in
spikes. 
 The spike-center distributions are obtained to exhibit the structure
expected from coherent gluon-jet emission dynamics. 
 This structure is similar in both cases considered, namely for all
charged and negatively charged particles, and is also similar to that
observed recently for all-charged-particle spikes in hadronic
interactions. 
 The energy distribution within spikes is found to have a significant
peak over the inclusive background, while the inclusive spectrum shows
exponential decrease with two characteristic values of average kinetic
energy.  
  The value of the peak energy and its width are in a good agreement with
those expected for pions produced in a nuclear medium in the framework of
the {\v C}erenkov quantum approach. 
 The peak energy obtained is consistent with the value of the
cross-section maximum observed in coincidence nucleon-nucleus interaction
experiments. 
 \end{abstract}

\maketitle



\section{Introduction}

A coherent component of particle production mechanism, {\v C}erenkov-like
radiation, in high energy particle collisions has been introduced a long 
time ago. The idea of mesonic \ct{ivan} and scalar (pionic) \ct{glas}
radiation has recently been a subject of the systematic analysis of
production
of mesons in high-energy pion-nucleon scattering in nuclear medium in
terms of classical quantum mechanics \ct{ions}. Characteristic signatures
of the {\v C}erenkov mechanism, such as the differential cross-sections
and  angle-energy correlations of produced particles, have been
predicted.

Another approach of the {\v C}erenkov-like radiation in strong
interactions was suggested within the QCD based coherent gluon-jet
emission model \ct{revd}. In this model the pseudorapidity distributions
of
centers of particle dense groups, called spikes,
are proposed to be investigated. The distributions of spike centers are   
predicted to have two peaks due to destructive interference for quarks of
the same colour (pp collisions) or to be singly peaked due to
constructive
interference for quarks of different colour (e.g. p$\bar {\rm p}$
interactions).  Recent observations in hadronic \ct{hh}
interactions are found to be in agreement with these
predictions.

The aim of this report is to present our recent results on
investigations of spikes in relativistic nucleus-nucleus collisions
\ct{pl98,pl99,yaf2k}. 
 We search for a coherent component of hadroproduction mechanism studying
two types of distributions assigned to the two above descibed scenarios. 
 Respectively, on the one hand, we analyse the distributions of centers
of spikes in the frame of the gluon-jet emission model and, on the other
hand, we investigate energy distribution within spikes to find the
features predicted by the nuclear pionic {\v C}erenkov-like radiation
(NPICR)  approach. 
 
To note is  that spikes have been extensively
investigated last years using a stochastic picture of particle
production,
namely an intermittency phenomenon has been searched for and obtained
in all types of collisions \ct{revi}. In our studies \ct{we}, we also
found the intermittency effect leading to multifractality and, then, to a
suggestion of a
possible non-thermal phase transition during the cascading.  
 The latter observation have been confirmed in different reactions
\ct{wec}.

\section{Experimental details}
  
The results presented here are based on the experimental data obtained
after processing the pictures from the 2m Streamer Chamber SKM-200
\ct{skm} installed in a 0.8~T magnetic field.
 The chamber was irradiated at the Dubna JINR Synchrophasotron. 
 A beam of magnesium nuclei with momentum 4.3 $A$ GeV$/c$ was used to
collide with a magnesium target, while a 4.5 $A$ GeV$/c$ carbon beam
interacted with a copper target.
 A central collision trigger was used to start the Chamber if there
were no charged or neutral projectile fragments (momentum per nucleon
required to be greater than 3 GeV$/c$) emitted in a forward cone of
2.4$^{\circ}$. 
 A more detailed description of the set-up design and data reduction
procedure are given elsewhere \ct{skm,skm1,idjmp}.  
  Systematic errors related to the trigger effects, low-energy pion and
proton detection, the admixture of electrons, secondary interactions in
the target nucleus etc.  have been considered in detail earlier and the
total contribution is estimated to not exceed 3\% \ct{skm1,skm2}.

A total of 14218 Mg-Mg events and 663 ones of C-Cu collisions were found
to meet the above centrality criterion. 
 In the utilized Mg-Mg sample only negative charged particles (mainly
$\pi^-$ mesons with a portion of some 1\% kaons) have been studied, while
in the C-Cu sample all charged aprticles have been considered. 

  In  the Mg-Mg events the average measurement error in momentum
$\al\ve_p/p\ar $ was about 1.5\% and that in the production angle
determination was $\al\ve_{\vt}\ar\se 0.1^{\circ}$.  
  The particles were selected in the pseudorapidity
($\e=-\ln\;\tan\frac{1}{2}\vt$)  window of $\Delta \eta =0.4 -2.4$ (in
the laboratory frame), in which the angular measurement accuracy does not
exceed 0.01 in $\e$ units.
 The mean multiplicity of the selected pions in the Mg-Mg sample is
$6.70\pm 0.02$. 

 In the C-Cu sample the average measurement error in momentum 
$\al\ve_p/p\ar$
was  about 12$\%$, the error in the polar angle was
$\al\ve_{\vt}\ar\se 2^{\circ}$, and the angular
measurement accuracy does not exceed 0.1 in $\e$-units in the
pseudorapidity range of $\De= 0.2-2.6$. 
 In addition, particles with $p_{\rT}>1$ GeV/$c$ are excluded from the
investigation as far as no negative charged particles were observed with
such a transverse momentum.  Under the assumption of an equal number of
positive and negative pions, this cut was applied to eliminate the
contribution of protons

To overcome an influence of the shape of the pseudorapidity distribution
on the results, we  use the ``cumulative variable'',

\be
\ef(\e)\; =
\int_{\e _{\mn}}^{\e}
\r(\e ')\, {\rd} \e ' \,
\Bigg/
\int_{\e _{\mn}}^{\e _{\mx}} \r(\e ')\, {\rd} \e '\; ,
\la{nv}
\ee

\noindent
with the uniform spectrum $\ddf$ within the interval [0,1], as advocated 
in Ref. \ct{fl}. This transformation makes it possible to
compare results from different experiments.   

\begin{figure*}[!htb]
\bcr
\epsfysize=8.2cm
\epsffile[ 20 185 505 600]{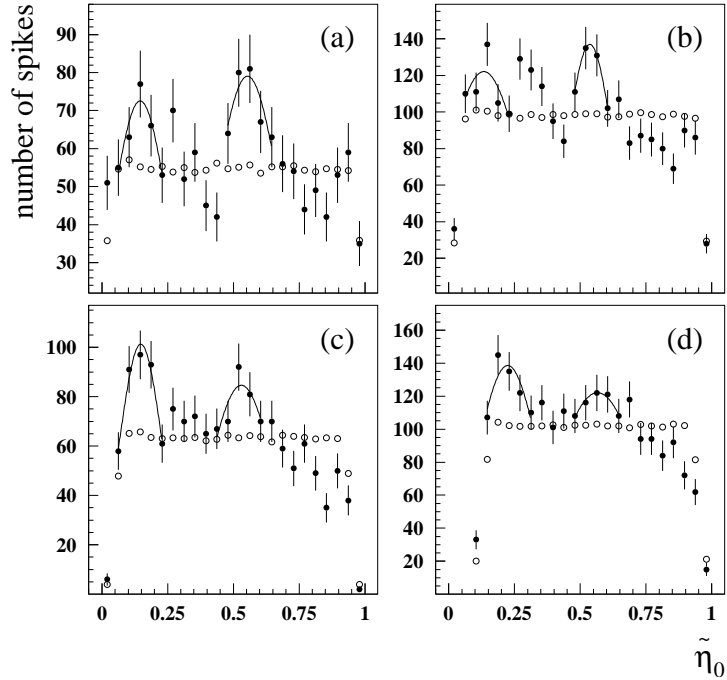}
\caption{Experimental ($\bullet$) and simulated ($\circ$)
spike-center distributions in C-Cu collisions
for  different $\dep$-bins
and
multiplicities $\dn$:
(a) $\dep\, = 0.04,\; \dn=4$,
(b) $\dep\, = 0.08,\; \dn=5$,
(c) $\dep\, = 0.12,\; \dn=7$,
(d) $\dep\, =  0.2,\; \dn=9$.
The curves represent Gaussian fits.
}
\vs{-.35cm}
\ecr
\la{cntrccu}
\end{figure*}

The spikes are extracted in each event from the ordered pseudorapidities
scanned with a fixed pseudorapidity window (bin) of size
$\dep$. Spikes with a definite number of particles $\dn$, hit in   
the bin, are determined and,
for each $\dn$,
the distribution of centers of spikes,
averaged over all events, is obtained. The center of spike is defined by
$\ecf\,=(1/\dn)\sum_{j=1}^{\dn}\ef_j$.

 To reveal dynamical correlations, the $\ecf$-distribution is
compared with analogous distributions obtained from the simulated
pseudorapidity single-particle spectrum $\ddf$ without any input
information about particle correlations.  The simulation procedure was as
follows.
 The number of particles was randomly generated
according to the multiplicity distribution of the data sample.
 Then, the pseudorapidities were distributed in accordance with the
experimental $\ef$-spectrum corresponding to the generated multiplicity.
 In each case of the reactions considered here, the total number of the
simulated events exceeded the experimental statistics by a factor of 100. 
 It is clear that the statistical properties of this set
are completely analogous to those of an ensemble resulting from arbitrary 
mixing of tracks from different events, subject to the condition of
retention of the $\ddf$-distribution. 
 So, the obtained sample represents independent particle emission.

\section{The results}

\subsection{Spike-center distributions}
\smallskip

The pseudorapidity spike-center $\ecf$-distributions for four
different-size $\dep$-bins and for spikes of various multiplicities $\dn$
are shown in Figs. \ref{cntrccu} and \ref{cntrmg} for C-Cu and Mg-Mg
collisions,
respectively.

For each reaction type, a two-peak structure of the measured
distributions (solid circles) is seen with the peaks in the neighbourhood
of the same $\ecf$, independent of the width and multiplicity of spike.
 The shape of the distributions is in agreement with the structure
predicted by the coherent gluon-jet emission model \ct{revd} and is
similar to that observed in hadronic interactions \ct{hh}.

In order to estimate the position of the peaks and the distance between
them, we fit these bumps with Gaussians and average over different
spikes.
 The peaks are found to be placed at $ \ecf\, \approx 0.17$ and $0.57$
corresponding to $\e_0=0.60\pm0.05\st\pm0.12\sy$ and
$1.30\pm0.03\st\pm0.10\sy$ in C-Cu collisions, and in
$\ecf\, \approx 0.19$ and 0.63 corresponding to $\e_0=
0.89\pm0.03\st\pm0.08\sy$ and $1.63\pm0.05\st\pm0.10\sy$ in Mg-Mg
interactions. 

\begin{figure*}[!t]
\bcr
\epsfysize=7.5cm
\epsffile[ 25 350 565 690]{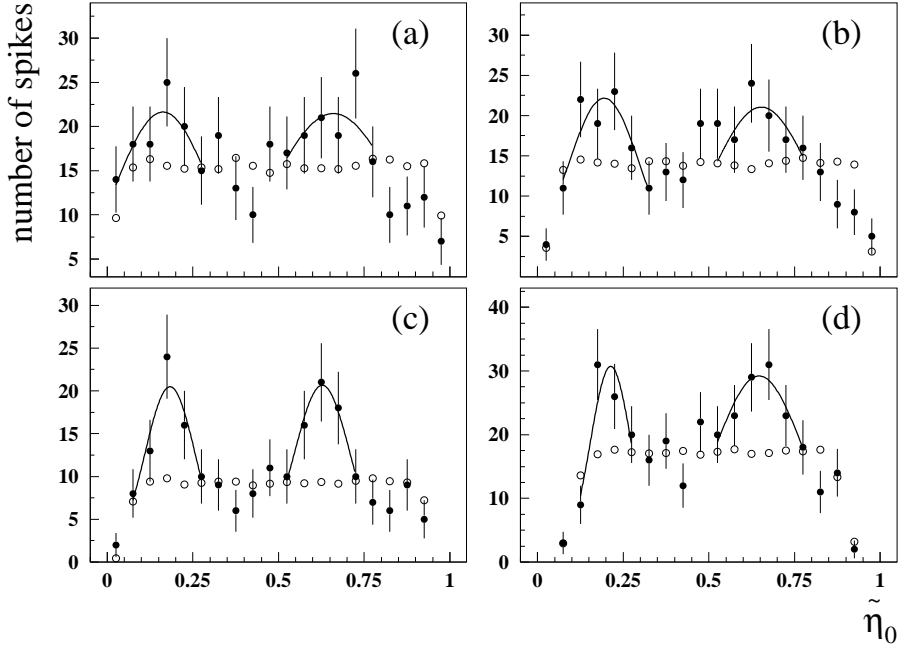}
\vs{.2cm}
\caption{Experimental ($\bullet$) and simulated ($\circ$)
spike-center distributions in Mg-Mg collisions 
for  different $\dep$-bins
and
multiplicities $\dn$:
(a) $\dep\, = 0.05,\; \dn=4$, (b) $\dep\, = 0.1,\; \dn=5$,
(c) $\dep\, = 0.15,\; \dn=6$, (d) $\dep\, =  0.25,\; \dn=7$.
The curves represent Gaussian fits.
}
\la{cntrmg}
\ecr
\vs{-.5cm}
\end{figure*}

They are separated by the following $d_0$ interval,
$$
{\hs{-.8cm} 
d_0=0.68\pm0.06\st\pm0.16\sy \;\:\; ({\mbox{\rm C-Cu}})}
$$
\be
d_0=0.75\pm0.06\st\pm0.13\sy \;\:\; ({\mbox{\rm Mg-Mg}})
\nonumber
\la{d}
\ee
in $\e$ units.
These values are close to those from the above mentioned hadronic
interactions.

The dynamical origin of the structure obtained is seen from a comparison
of
the experimental $\ecf$-distributions with those based on the
above described simulated
events (open circles in Figs. \ref{cntrccu} and \ref{cntrmg}). No peaks
are seen in the
simulated
distributions, levelling off at the background, far below the
measured
peaks. This points to a dynamical effect in the formation of spikes
in agreement with
the coherent gluon radiation picture.

\begin{figure*}[!htb]
\bcr
\epsfysize=7.9cm
\epsffile[5 14 615 485]{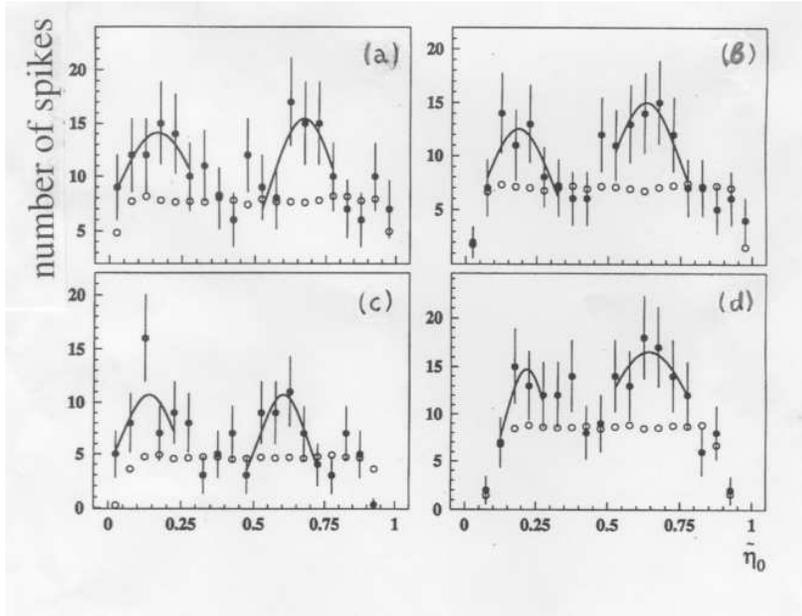}
\vs{-.7cm}
\caption{The same as in Fig. \ref{cntrmg} but for 
azimuthally isotropic events (see text and Eq. (\ref{beta})).
}
\vs{-.4cm}
\la{cntrmgj}
\ecr
\end{figure*}

In order to reveal a possible contribution of hadronic jets to the effect
observed, for negative pions from Mg-Mg collisions we carried out
additional analysis taking into account
azimuthal collinearity. To this end, the alignment coefficients,

\be
\beta=\frac{\sum_{i\neq j}^{n}\cos 2(\Phi_i-\Phi_j)}{\sqrt{n(n-1)}}\, ,
\la{beta}
\ee
 for an individual event with multiplicity $n$ were used with 
$\Phi_i$ being the $i$th particle azimuthal angle \ct{dreman}.  

A $\beta<0$ criterion were
applied to minimize  a contribution of the jet-structural events.
This reduced statistics twice. The resulted  $\ecf$-distributions are
shown 
in Fig. \ref{cntrmgj} at the same  $\dep$ and $\dn$ as in Fig.
\ref{cntrmg}
without the $\beta$-criterion.   
One can see that the structure of the $\ecf$-distributions for such
azimuthally isotropic events 
does not change
compared to that of Fig. \ref{cntrmg} and the distributions demonstrate
two peaks. The Gaussian fits of the peaks give the peaks positions
to be 
$\ecf\, \approx 0.88$ and 1.63 with the distance between them
$d_0=0.75\pm0.06\st\pm0.15\sy$. The obtained values are almost the same 
as those found from Fig. \ref{cntrmg}.

The nearness of the positions of the peaks with and without
$\beta$-criterion points at the azimuthal isotropy of events with spikes.

In order to assess the reliability of the results obtained, we studied
the influence of the $\De$-range used and the experimental error
$\al\ve_{\vt}\ar$ in the measurements of the polar angle $\vt$.
 Varying the $\De$ range and the error $\al\ve_{\vt}\ar$ we found the
structure of the distributions unchanged and the positions of the two
peaks and the distance $d_0$ to be within the above shown systematic
errors, in support of the conclusions made. 

\subsection{In-spike energy spectra}
\smallskip
\smallskip

The strong signal of the coherent emission dynamics obtained allows
further search for its manifestation in the energy distribution, as
predicted in the NPICR approach \ct{ions}. In this model, the energy   
spectrum of pions, emitted through the coherent {\v C}erenkov-like
mechanism
when a few-GeV proton passes the nuclear medium, is predicted to have a
peak. This peak is expected to appear at 260 MeV when an absorption
effect
is 
neglected and at 244 MeV otherwise.

In Fig. \ref{kinen} we compare the c.m.s. inclusive kinetic energy
distribution,
$F(K^*)$ $=(1/E^*p^*) \,{\rd}N/{\rd}K^*$, with analogous spectra
calculated for pions from
spikes in Mg-Mg interactions. Here, $E^*$ and $p^*$ denote, respectively,
the particle energy
and momentum in the c.m.s. frame.

\begin{figure*}[!t]
\vs{1.1cm}
\epsfysize=6.5cm
\bcr
\epsffile[25 350 565 695]{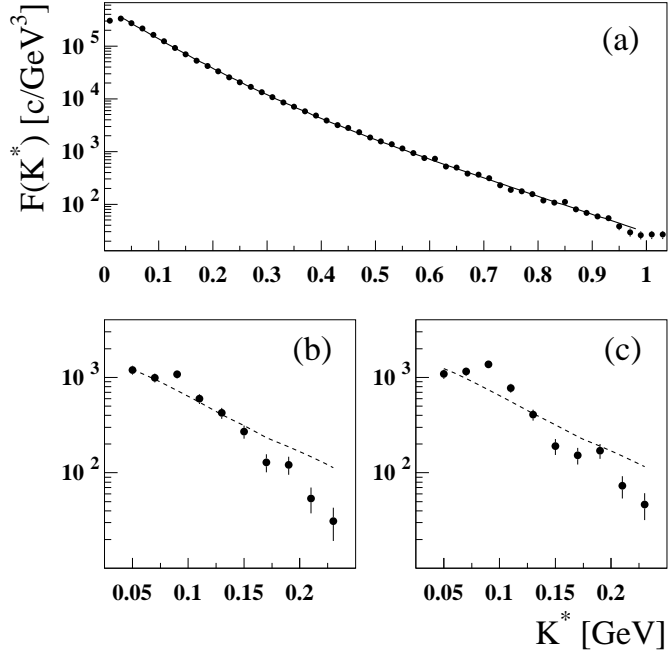}
\vs{.3cm}
\caption{Inclusive kinetic energy distribution (a)
        and
        analogous distributions for spikes of negative pions in Mg-Mg
interactions, 
        (b) $\dep\, = 0.1,\; \dn=6$ 
        and
        (c) $\dep\, = 0.15,\;\dn=7$.
The solid line represents the exponential fit with Eq. (\ref{teme}),   
the dashed lines show the inclusive background. 
}
\vs{-.5cm}
\la{kinen}
\ecr
\end{figure*}

Using the temperature description, utilized to
characterize a system of excited hadrons \ct{idjmp, broc, nagu, backo},
we
parametrize the inclusive spectrum (Fig. \ref{kinen}a) by a sum of two
exponents,

\be
F(K^*) = A_1 \exp(-K^*/T_1)+A_2 \exp(-K^*/T_2)\, ,
\la{teme}
\ee

\noindent
where the temperatures $T_{1}<T_{2}$ characterize \ct{broc} the two   
possible
mechanisms of pion production, via $\Delta$-resonance decay and directly,
and are related to the pion average kinetic energies. The range of the
parametrization
shown is limited from below and from above due to detector effects and
corresponding requirements on the momenta of pions. The fit gives $T_1=
65
\pm 1$ MeV
and $T_2= 127\pm 1$ MeV. These values are, in general, consistent with  
those obtained from the earlier analysis of the reaction under study
\ct{idjmp} and from other experiments \ct{broc,nagu}.  Some difference in
the values could be explained if one takes into account the difference in
the sizes of the (pseudo)rapidity regions used \ct{backo}.

The shape of the $F(K^*)$ distribution changes when the analysis is
extended
to spikes, Figs. \ref{kinen}b and \ref{kinen}c.  The energy spectrum of
particles
belonging
to a
spike differs significantly from the exponential law (\ref{teme}) and has 
a peaked shape. To extract the NPICR-signal we compare the
in-spike energy spectra with the renormalized inclusive distribution, or
inclusive background, depicted by the dashed lines.  The first peak is   
seen to be above background with a statistical significance of 2.7 and 4.1
standard deviations in Figs. \ref{kinen}b and \ref{kinen}c, respectively.
This peak is
located at the kinetic energy $K^* \approx 100$ MeV, or the total energy
$E^* \approx 240$ MeV, in accordance with the NPICR prediction.

To estimate the position of the peak and its width and to make the
results more comparable with the NPICR expectations, the
$E^*$-distributions of particles in spikes of various size $\dep$-bins
and different $\dn$-multiplicities are studied.  Fig. \ref{toten} 
represents
examples for
these distributions. The following specific peculiarities are found.

All these $E^*$-distributions possess a non-exponential behaviour with a
pronounced maximum in the vicinity of the value $E_{\rm m}^*=240$ MeV   
regardless of
bin size and multiplicity of spikes. The higher the multiplicity of spike
is (at fixed $\dep$ size), the more peaks appear.
A multi-peak structure is observed for bins with the multiplicities
$\dn>3$, while at $\dn \leq 3$ only one peak occurs (not shown).

\begin{figure*}[!htb]
\bcr
\epsfysize=8.cm
\epsffile[25 300 565 695]{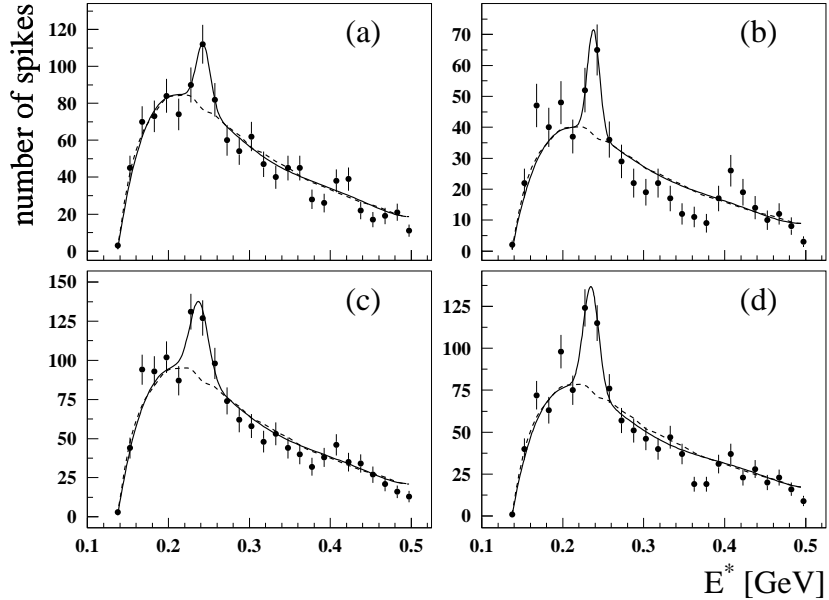}
\vs{-.9cm}
\caption{Total energy spectra for negative particles spikes of
        different $\dep$-bins and multiplicities $\dn$ in Mg-Mg
collisions:
(a) $\dep\, = 0.05,\; \dn=4$, (b) $\dep\, = 0.08,\; \dn=5$,
(c) $\dep\, = 0.1,\; \dn=5$,  (d) $\dep\, =  0.15,\; \dn=6$.
The solid lines represent the fit (see text), the dashed lines show the
inclusive background. 
}
\la{toten}
\ecr
\vs{-.6cm}
\end{figure*}

To reveal the dynamical signal we compare the in-spike energy
distributions
with the inclusive background (dashed-line). As for the above
kinetic-energy distributions, the value of $E^*$ of about 240 MeV is
obtained
to be the position of the
most prominently and statistical-significantly peak over background.
To estimate the background and to parametrize the signal, we use a
fifth-order polynomial for the background and a Gaussian for the peak.  
The solid curve shows the result of this fit.
After averaging over various spikes, the position of the peak and its
width are found to have the values,

$$
{\hs{-2.2cm} E^*_{\rm m}=238 \pm 3{\st}\pm 8{\sy} \; {\rm MeV},}
$$
\be
 \Gamma_{\rm m}=10 \pm 3{\st}\pm 5{\sy} \; {\rm MeV},
\la{eg}
\ee

\noindent
respectively.

The location
of the obtained centre of the Gaussian lies within the interval of   
energies expected for pions from the {\v C}erenkov-like mechanism for
incident protons of a
few GeV, $224\leq E_{\rm m}\leq244$ MeV \ct{ions}. The value of   
$E^*_{\rm m}$
(\ref{eg}) is similar to the position of the peak observed in the
$\pi^+$p invariant mass distribution in the analysis of coincidence
measurements of (p,n) reactions on carbon at 1.5 GeV/$c$ in the
$\Delta$-resonance excitation region \ct{chib}, the effect connected with
the NPICR mechanism \ct{ions}.
Also, the width $\Gamma_{\rm m}$ confirms an observation of the {\v
C}erenkov
radiation signal expected to be  $\Gamma \leq 25$ MeV.

\section{Conclusions}

In summary, in order to search for a coherent, {\v C}erenkov-like
emission mechanism of particle production, a study of spikes in
relativistic nuclear collisions is carried out with charged particles from
central C-Cu collisions at a momentum of 4.5 $A$ Gev/$c$ and with  
negative pions from
central Mg-Mg collisions at a momentum of 4.3 GeV/$c$ per incident
nucleon.
The spike-center distributions and the energy
spectra
of particles within  a spike are investigated for various
narrow pseudorapidity bins and different spike multiplicities.

The spike-center distributions are found to possess a double-peak shape
that is in agreement with the structure expected from the coherent gluon
radiation model.
The obtained distance between the peaks as well as  the shape of the 
distributions are similar to those observed recently in analogous studies
of charged-particle spikes in hadronic
interactions.
The dynamical effect in the spike-center distributions is revealed in a
comparison with an independent particle-emission model, where no peaks
are seen.

The coherent character of the particle-production mechanism is confirmed
by
studying energy distributions in Mg-Mg interactions.  The inclusive energy
spectra show
monotonic exponential decrease with two specific temperatures, while the 
in-spike energy distributions are obtained to exhibit a peak at a
position
and of a width both consistent with the values expected from
the theoretical
calculations based on the
hypothesis of nuclear pionic {\v C}erenkov radiation. The value  of the
peak energy is close to the recently observed maximum in the
differential
cross-sections studied in the coincidence experiments of a few GeV
nucleon-nucleus interactions.

The results of the presented analysis signalize coherent emission
as a complementary mechanism to the stochastic scenario of
hadroproduction. Furthermore, the similarity of the
spike-center distributions obtained for like-charged particles in the
presented paper with those for all-charged-particle spikes in the earlier
studies
indicates important contributions of the coherent mechanism
to the formation of Bose-Einstein correlations \ct{bec}. It is worth to
mention that, in comparison to stochastic (intermittency) dynamics,
the origin of which remains still unclear \ct{revi}, the coherent
emission
has definite underlying dynamics.
All this gives evidence for the necessity of further efforts
in studying existing experimental data.

\section*{\small\bf Acknowledgements}

One of the authors (E.S.) would like to thank the organisers of the
Workshop on
Multiparticle Production for their very kind hospitality
and financial support.


\end{document}